\documentclass[referee,a4paper]{raa}           

\usepackage{graphicx,booktabs,threeparttable}
\usepackage{natbib}
\usepackage{amssymb,amsmath}
\usepackage[pagebackref=true]{hyperref}

\begin{document}
\title{Injection spectra of different species of cosmic rays from AMS-02, ACE-CRIS and Voyager-1}
\titlerunning{Injection spectra of different CR species}

\volnopage{ {\bf 20XX} Vol.\ {\bf X} No. {\bf XX}, 000--000}
   \setcounter{page}{1}
\author{Xu Pan\inst{1,2}, Qiang Yuan\inst{1,2}}


\institute{Key Laboratory of Dark Matter and Space Astronomy, Purple Mountain Observatory, Chinese Academy of Sciences, Nanjing 210023, P. R. China; {\it yuanq@pmo.ac.cn}\\
\and  
School of Astronomy and Space Science, University of Science and Technology of China, Hefei 230026, P. R. China}

\abstract{
Precise measurements of energy spectra of different cosmic ray species 
were obtained in recent years, by particularly the AMS-02 experiment on 
the International Space Station. It has been shown that apparent differences 
exist in different groups of the primary cosmic rays. However, it is not 
straightforward to conclude that the source spectra of different particle 
groups are different since they will experience different propagation 
processes (e.g., energy losses and fragmentations) either. In this work, 
we study the injection spectra of different nuclear species using the
measurements from Voyager-1 outside the solar system, and ACR-CRIS and
AMS-02 on top of atmosphere, in a physical framework of cosmic ray
transportation. Two types of injection spectra are assumed,
the broken power-law and the non-parametric spline interpolation form.
The non-parametric form fits the data better than the broken power-law form,
implying that potential structures beyond the constrained spectral shape
of broken power-law may exist. For different nuclei the injection spectra are 
overall similar in shape but do show some differences among each other. 
For the non-parametric spectral form, the helium injection spectrum is 
the softest at low energies and the hardest at high energies. 
For both spectral shapes, the low-energy injection spectrum of neon is the 
hardest among all these species, and the carbon and oxygen spectra have more 
prominent bumps in $1-10$ GV in the $R^2dN/dR$ presentation.
Such differences suggest the existence of differences in the sources or
acceleration processes of various nuclei of cosmic rays.
}

\maketitle

\section{Introduction}
Although the exact origin of cosmic rays (CRs) is not clear yet, it is 
generally believed that CRs with energies below PeV originate from supernova 
remnants. Energetic CRs were accelerated by diffusive shocks and then 
injected into the interstellar space. Theoretically the accelerated 
spectrum can be simply described by a power-law form $dN/dR\propto R^{-n}$, 
with $R$ being the particle rigidity and $n$ being the index
\citep{1949PhRv...75.1169F,1978MNRAS.182..147B,1978ApJ...221L..29B,2014BrJPh..44..415B}. 
Extension of the conventional diffusive shock acceleration mechanism with 
test particle assumption to consider the interaction between accelerated 
particles and the surrounding fluid results in nonlinear effects and
deviation from the simple power-law spectrum
\citep{2001RPPh...64..429M,2004MNRAS.353..550B,2010APh....33..307C}.
From the observational point of view, complicated spectral structures of
CRs were also revealed by many measurements 
\citep{2009BRASP..73..564P,2010ApJ...714L..89A,2011Sci...332...69A,
2015PhRvL.114q1103A,2015PhRvL.115u1101A,2017PhRvL.119y1101A,
2018JETPL.108....5A,2019SciA....5.3793A,2019PhRvL.122r1102A,
2020PhRvL.125y1102A,2021PhRvL.126t1102A,2022DAMPE-BC}. 
Particularly, apart from the breaks around a few GV, remarkable hardenings 
around hundreds of GV and subsequent softenings around 10 TV were shown by 
the data. The spectra also differs among different nuclei. The helium 
spectrum is found to be clearly harder than that of protons
\citep{2011Sci...332...69A,2015PhRvL.114q1103A,2015PhRvL.115u1101A}.
The AMS-02 measurements further showed that the high-energy spectra
of neon (Ne), magnesium (Mg), and silicon (Si) are different from those
of helium (He), carbon (C), and oxygen (O), and suggested that different
types of primary sources exist \citep{2020PhRvL.124u1102A}. These results
may indicate that the origin and acceleration of CRs are more complicated.

It should be noted that after the acceleration, CR particles are injected
into the interstellar space, and experience complex propagation processes.
The energy losses and fragmentation cross sections of various nuclei differ 
from each other, making the propagated spectra become diverse even for
the same injection spectra. Therefore, the apparent differences of the 
spectra are not directly reflecting the differences at injection. To properly 
address this issue needs a thorough consideration of the CR propagation
\citep{2017PhRvD..95h3007Y,2018ApJ...858...61B,2019A&A...627A.158D,
2020ApJ...889..167B,2020ApJS..250...27B,2019SCPMA..6249511Y,
2019PhLB..789..292W,2022ApJ...932...37N,2022PhRvD.105j3033K}.

Here we investigate the source injection spectra of different primary
nuclei including He, C, O, Ne, Mg, Si, and Fe, based mainly on the AMS-02
data \citep{2017PhRvL.119y1101A,2020PhRvL.124u1102A,2021PhRvL.126d1104A,
2021PhR...894....1A}. 
At low energies the fluxes will be suppressed due to the solar modulation
effect. We use the force-field approximation to account for the solar
modulation \citep{1967ApJ...149L.115G}. To break the degeneracy between
the injection and the solar modulation effects, the measurements at 
low energies outside the solar system by Voyager-1 will also be included
\citep{2016ApJ...831...18C}. We further use the 
ACE-CRIS \footnote{\url{http://www.srl.caltech.edu/ACE/ASC/level2/lvl2DATA_CRIS.html}}
measurements at the same time periods of the AMS-02 to better constrain 
the low-energy spectra. The GALPROP code is employed to calculate the 
propagation of CRs \citep{1998ApJ...509..212S,1998ApJ...493..694M}.
The Markov Chain Monte Carlo (MCMC) method is used to do the fit
\citep{2012PhRvD..85d3507L}. 

Compared with previous works along the line of studying the
injection spectra of CRs 
\citep{2019SCPMA..6249511Y,2022PhRvD.105j3033K,2022ApJ...932...37N}, 
this work differs in either more species of nuclei used (e.g., Ne, Mg, Si, Fe)
or the low-energy ACE and Voyager data included which better constrain the 
wide-band spectral shape.

\section{Cosmic ray injection and propagation}

Given more and more complicated structures of the CR spectra were revealed by 
recent precise measurements, it is expected that simple empirical functions
may not be proper enough to describe the injection spectra of CRs in a wide 
energy range. In this work, we use a non-parametric interpolation (NPI) spectrum 
determined by a cubic spline interpolation method \citep{2016A&A...591A..94G,
2018ApJ...863..119Z}, which has more freedom to reveal multiple structures
of the spectra. The interpolation is done in the $\log(R)-\log(J)$ parameter 
space, where $R=pc/Ze$ is the particle rigidity in unit of MV and $J$ is
the flux. Specifically, we set the following rigidity knots in the analysis:
\begin{equation}
\left\{ \log(R_{1}),..., \log(R_{7}) \right\} = \left\{ 2.50,3.11,3.72,
4.29,4.91,5.49,6.10 \right\}.
\end{equation}
The corresponding fluxes at these rigidity knots, $\log(J_i)$, are fitted
as free parameters.

To capture the main features of the spectra, we also consider
a broken power-law (BPL) form of injection spectrum for comparison
\begin{equation}
q(R) \propto \begin{cases}\left(R / R_{\mathrm{br0}}\right)^{-\gamma_0}\left(R_\mathrm{br0}/R_\mathrm{br1}
\right)^{-\gamma_1}, & R\leq R_{\mathrm{br0}} \\  \left(R / R_{\mathrm{br1}}\right)^{-\gamma_1}, &
R_{\mathrm{br0}} <R\leq R_{\mathrm{br1}} ,\\ \left(R / R_{\mathrm{br1}}\right)^{-\gamma_2}, & R \geq R_{\mathrm{br1}}\end{cases}
\end{equation}
where $\gamma_0$, $\gamma_1$, and $\gamma_2$ are spectral indices in 
different rigidity ranges, $R_{\mathrm{br0}}$ and $R_{\mathrm{br1}}$ are break rigidities.

Following the distribution of supernova remnants, the source distribution 
of CRs is parameterized as
\begin{equation}
f(r, z)=\left(\frac{r}{r_{\odot}}\right)^\alpha \exp \left[-\frac{\beta\left(r-r_{\odot}\right)}{r_{\odot}}\right] \exp \left(-\frac{|z|}{z_s}\right),
\end{equation}
where $r_{\odot}=8.5$ kpc, $z_s=0.2$ kpc, $\alpha=1.25$, $\beta=3.56$ \citep{2011ApJ...729..106T}.

The propagation of nuclei in the Milky Way includes mainly the diffusion in
the random magnetic field, the energy losses due to ionization and Coulomb
collisions, the fragmentation due to inelastic collisions with the 
interstellar medium (ISM), and possible convective transportation and
reacceleration \citep{1964ocr..book.....G,2007ARNPS..57..285S}.
The propagation can be described by a set of differential equations
for all species of nuclei, which self-consistently predict the fluxes
of both primary and secondary nuclei. The general propagation equations
can not be solved analytically, and numerical solutions were developed
and widely employed \citep{1998ApJ...509..212S,1998ApJ...493..694M}.

The propagation parameters we adopt are determined through fitting to the
newest measurements of secondary and primary CRs \citep{2020JCAP...11..027Y}.
We work in the diffusion-reacceleration framework, and the convection velocity 
is set to be 0. The main parameters include: the spatial diffusion coefficient 
$D_{xx}=D_{0}\beta^{\eta}(R/4~{\rm GV})^{\delta}$, with $D_0=7.69\times10^{28}$
cm$^2$~s$^{-1}$, $\eta=-0.05$ which phenomenologically describes the
possible resonant interactions of CRs with the magnetohydrodynamic (MHD) 
waves \citep{2006ApJ...642..902P}, $\delta=0.362$, the Alfvenic velocity
$v_A=33.76$ km s$^{-1}$ which characterizes the reacceleration of particles 
during the propagation, and the half height of the propagation halo 
$z_h=6.27$ kpc.

After entering the solar system, CRs would be further affected by the 
magnetic field carried by the solar wind, and experience flux suppression
at low energies (below tens GV). This so-called solar modulation results 
in an anti-correlation of the low-energy CR fluxes with solar activities. 
Although more sophisticated modulation models were developed 
(e.g., \cite{2013LRSP...10....3P}), the simple force-field approximation 
\citep{1967ApJ...149L.115G} is employed in this work. Since the particles
discussed here are all positively charged with mass-to-charge ratio 
$A/Z\approx2$, we expect that their relatively differences are less
sensitive to the solar modulation model.

\section{Analysis method}
In this work we focus on the primary CR nuclei with $A/Z \approx 2$,
including He, C, O, Ne, Mg, Si, and Fe. The proton spectrum which shows
clear difference from that of He is not discussed \citep{2017ApJ...844L...3Z}. 
The CosRayMC code \citep{2012PhRvD..85d3507L} which combines the CR propagation
and the MCMC sampler is used. According to the Bayes' theorem, the posterior 
probability of a model described by parameters $\boldsymbol{\theta}$ can be
obtained as
\begin{equation}
P(\boldsymbol{\theta}|D) \propto {\mathcal L}(\boldsymbol{\theta})
P(\boldsymbol{\theta}),
\end{equation}
where $D$ denotes the data, ${\mathcal L}(\boldsymbol{\theta}) = 
P(D|\boldsymbol{\theta})$ is the likelihood of the model for data $D$,
and $P(\boldsymbol{\theta})$ is the {\it prior} probability of the model.
The likelihood function can be calculated as 
\begin{equation}
{\mathcal L}(\boldsymbol{\theta})\propto\exp\left(-\frac{1}{2}{\sum_{i=1}^{n}
\frac{[J(E_{i};\boldsymbol{\theta})-J_i]}{\sigma_{i}^{2}}}\right),
\end{equation}
where $J(E_{i};\boldsymbol{\theta})$ is the model predicted flux, 
$J_i$ and $\sigma_{i}$ are the observational flux and error of the $i$-th
energy bin. 

The AMS-02 and Voyager-1 data can be directly obtained from the publications
\citep{2021PhR...894....1A,2016ApJ...831...18C}. The total uncertainties used
are the quadratic sum of the statistical ones and systematic ones. For the 
ACE-CRIS data, we extract them from the online data server. The systematic 
uncertainties of ACE-CRIS data include the geometry factor (2\%), the 
scintillating optical fiber trajectory efficiency (2\%), and the correction 
of spallation in the instrument ($1\% \sim 5\%$ depending on the charge and 
energy bin) \citep{2009ApJ...698.1666G}. For He nuclei, no ACE-CRIS data are
available. For Fe nuclei, the ACE-CRIS data are not included in the likelihood
calculation due to the possible excess compared with the AMS-02 data
(see the discussion in \cite{2021ApJ...913....5B}).

\section{Results}

\begin{figure}
\centering
\includegraphics[width=0.48\textwidth]{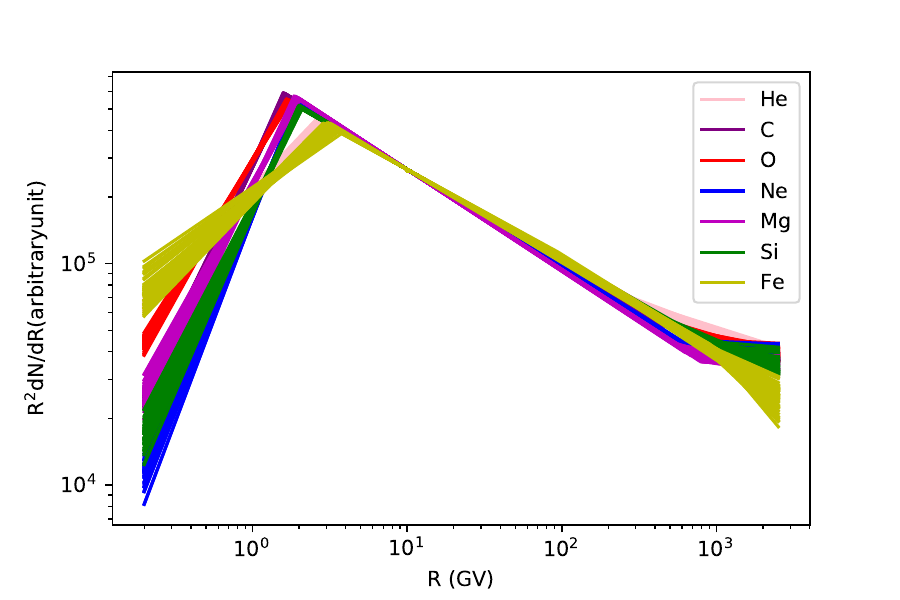}
\includegraphics[width=0.48\textwidth]{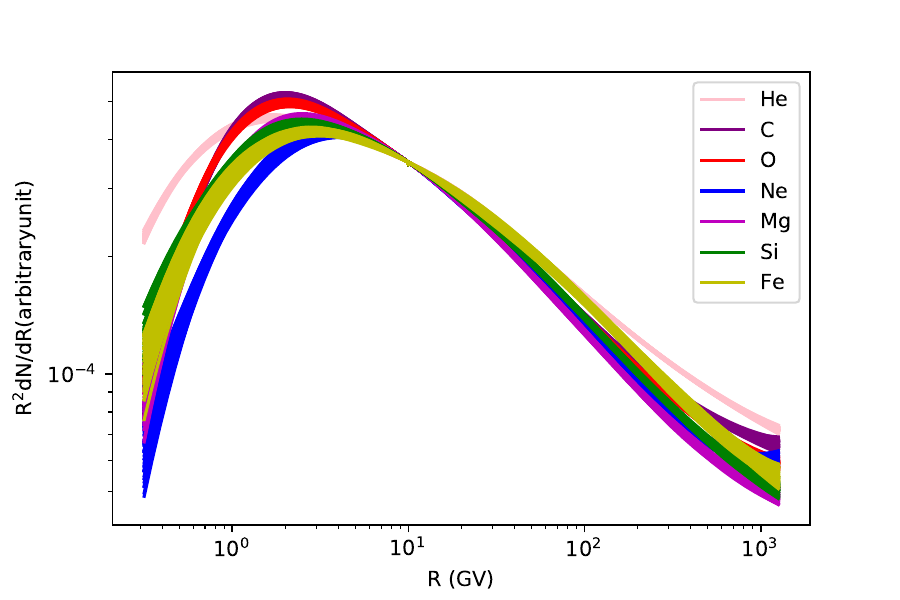}
\includegraphics[width=0.48\textwidth]{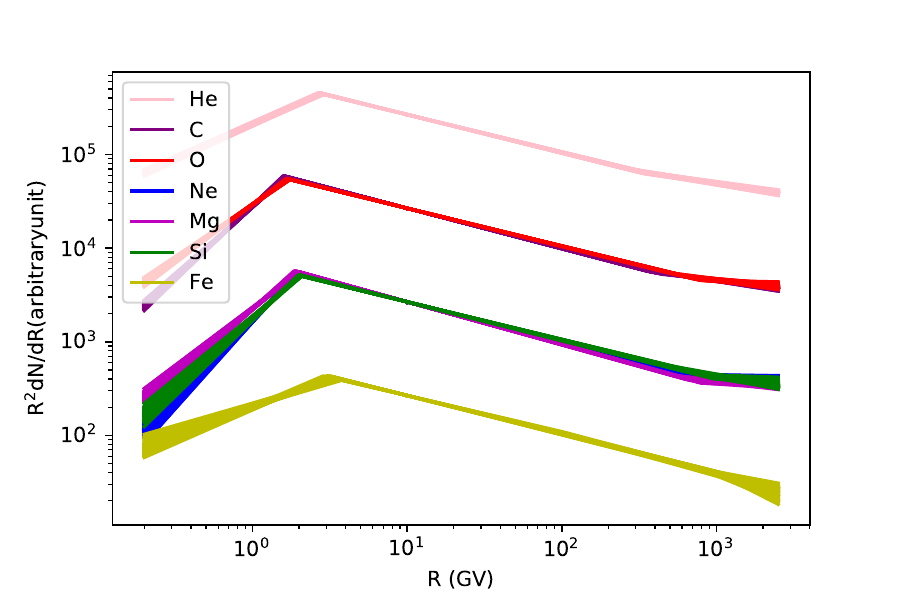}
\includegraphics[width=0.48\textwidth]{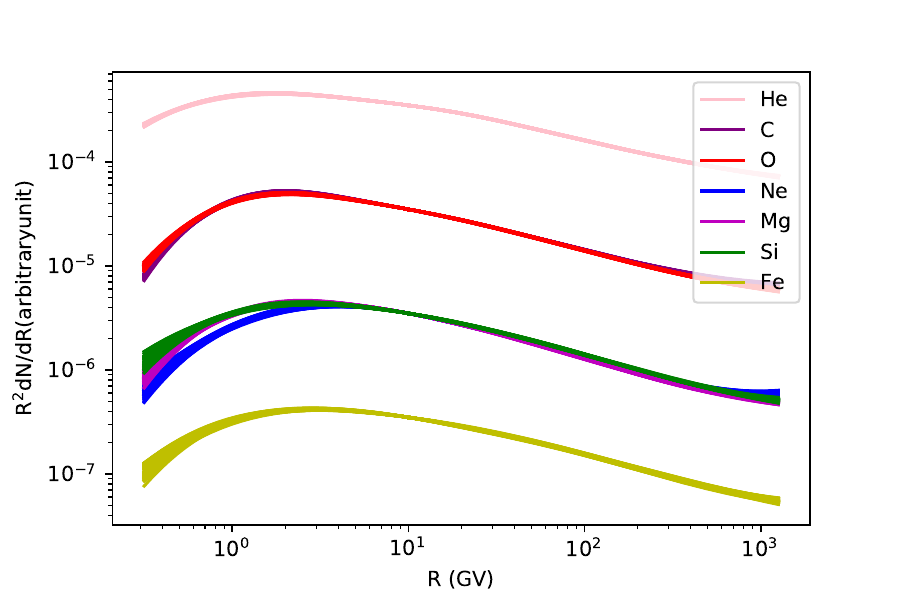}
\caption{The injection spectra of different nuclei. In the top panels we normalize all spectra at 10 GV, and in the bottom panels they are shown for 4 different groups. The left panels are for the BPL form, and the right panels are for the NPI form.}
\label{fig:injection}
\end{figure}

\begin{table}[ht]
    \centering
    \caption{The parameters of BPL form of injection spectra.}
    \scalebox{0.85}{
\begin{tabular}{cccccccc}
    \toprule
      &He &C &O&Ne &Mg & Si& Fe \\
    \midrule
    $\gamma_0$& $1.27^{+0.02}_{-0.02}$ & $0.50^{+0.09}_{-0.10}$ &$0.82^{+0.07}_{-0.08}$ & $0.30^{+0.14}_{-0.20} $&$0.62^{+0.12}_{-0.16} $ &$0.56^{+0.11}_{-0.11}$ &$1.37^{+0.11}_{-0.12}$ \\
    $\gamma_1$& $2.41^{+0.01}_{-0.01}$& $2.43^{+0.00}_{-0.00}$ &$2.41^{+0.00}_{-0.00}$ & $2.44^{+0.01}_{-0.01}$&$ 2.45^{+0.01}_{-0.01}$ &$2.41^{+0.01}_{-0.01} $ &$2.41^{+0.01}_{-0.01}$ \\
    $\gamma_2$& $2.25^{+0.02}_{-0.03}$ & $2.21^{+0.04}_{-0.06} $ &$2.14^{+0.09}_{-0.08}$ & $2.14^{+0.11}_{-0.09}$&$ 2.13^{+0.12}_{-0.09}$ &$2.27^{+0.01}_{-0.01}$ &$2.46^{+0.24}_{-0.16}$ \\
    $R_{\rm br0}$\:(GV) & $2.78^{+0.07}_{-0.07}$ & $1.63^{+0.05}_{-0.06}$ &$1.71^{+0.06}_{-0.06}$ & $1.94^{+0.40}_{-0.39}$&$1.93^{+0.08}_{-0.08} $ &$2.03^{+0.06}_{-0.04}$ &$3.29^{+0.39}_{-0.29}$ \\
    $R_{\rm br1}$\:(GV)& $306.30^{+36.22}_{-34.78}$ & $ 389.72^{+73.95}_{-59.53}$ &$692.38^{+140.72}_{-154.58}$ & $669.79^{+159.01}_{-143.47}$&$ 667.32^{+162.19}_{-152.56}$ &$791.60^{+138.20}_{-207.51}$ &$705.66^{+199.56}_{-314.79}$ \\
    $\phi$ & $0.70^{+0.01}_{-0.01} $ & $0.71^{+0.01}_{-0.01}$ & $0.67^{+0.01}_{-0.01} $& $0.76^{+0.02}_{-0.02}$&$0.70^{+0.01}_{-0.01}$ & $0.72^{+0.02}_{-0.02}$ & $1.00^{+0.04}_{-0.04} $             \\
    $\chi^2/\mathrm{d.o.f.}$ & 91.3/77 &60.0/80 & 54.3/81 & 81.2/76 & 80.3/76& 69.6/76 & 24.1/48 \\
      
    \bottomrule   

   \end{tabular}
   }
    \label{tab:bpl}
\end{table}

\begin{table}[ht]
    \centering
    \caption{The parameters of NPI form of injection spectra.}
    \scalebox{0.8}{
    \begin{threeparttable}
\begin{tabular}{cccccccc}
    \toprule
      &He &C &O&Ne &Mg & Si& Fe \\
    \midrule
    $\log(J_1)$& $-2.65\pm 0.02 $ & $-3.11\pm 0.05$ &$-3.01\pm 0.05$ & $-3.02\pm0.10 $&$-2.99\pm 0.10 $ &$-2.86\pm 0.10$ &$-2.89\pm0.12$ \\
    $\log(J_2)$\tnote{1}&  $-3.5$ & $-3.5$ &$-3.5$ &  $-3.5$ & $-3.5$ &$-3.5$ &$-3.5$\\
    $\log(J_3)$& $ -4.80\pm 0.01 $ & $-4.78\pm 0.01 $ & $-4.77 \pm 0.01 $ & $-4.61 \pm 0.02$ &$-4.70\pm 0.01$ &$ -4.72\pm0.02 $ &$-4.69\pm0.02$ \\
    $\log(J_4)$& $-6.14 \pm 0.01 $ & $ -6.17 \pm 0.01$ &$-6.16\pm 0.01$& $ -5.99\pm 0.02 $&$-6.09\pm0.02 $ & $-6.08\pm 0.02$ &$-6.04\pm0.02 $ \\
    $\log(J_5)$ & $-7.56\pm 0.01  $ & $ -7.62 \pm 0.02 $ &$-7.62 \pm 0.01 $ & $-7.44\pm 0.02 $ &$ -7.57\pm 0.02$ &$-7.54\pm 0.02$ &$-7.46\pm0.02$ \\
    $\log(J_6)$ & $-8.97 \pm 0.01 $ & $-9.0
    6 \pm 0.02  $ &$-9.07 \pm 0.01 $ & $-8.92\pm 0.02$&$-9.05\pm0.02$ &$ -9.02\pm 0.02$ &$-8.93\pm0.02$ \\
    $\log(J_7)$ & $-10.35 \pm 0.02 $ & $ -10.39\pm 0.02 $ &$-10.44\pm 0.01$ & $-10.25\pm 0.03 $&$-10.43\pm 0.03$ &$-10.43\pm 0.03$ &$-10.36\pm0.05$ \\
    
    $\phi$& $0.56\pm 0.01$ & $0.60\pm0.01$ &$0.56\pm 0.02$ & $0.51\pm 0.02 $&$ 0.50\pm 0.02$ &$0.50\pm 0.02$ &$0.69\pm 0.05$ \\
    
    $\chi^2/\mathrm{d.o.f.}$ &45.2/75 & 24.7/78 &  14.2/79 & 20.0/74 &24.2/74 & 29.5/74 &  13.3/46   \\

    \bottomrule   
\end{tabular}
  \begin{tablenotes}
        \footnotesize
        \item[1] The $\log(J_2)$ of all nuclei is normalized to the same constant value.
      
      \end{tablenotes}
   \end{threeparttable}        
}    
    \label{tab:non}
\end{table}

The best-fit parameters and the $1\sigma$ uncertainties for the BPL and NPI injection spectra are presented in Tables 1 and 2. For all species, the NPI form show smaller reduced $\chi^2$ values compared with those of the BPL form. Figure~\ref{fig:injection} displays the injection spectra of various nuclei (the $1\sigma$ bands) obtained from the fitting. Top panels are for the injection spectra normalized at 10 GV, and bottom panels display the spectra which are grouped into 4 groups, He, C-O, Ne-Mg-Si, and Fe, respectively.These injection spectra show a general similarity among each other. Specifically, the injection spectra for all nuclei experience softenings around several GV rigidities and hardenings around a few hundred GV. For the BPL form, our results of C, O, Ne, and Mg are consistent with those given in \cite{2022ApJ...932...37N}, despite the methodologies are different. The relative spectral shapes among different nuclei are different for the BPL and NPI forms. Since the NPI form introduces less constraints on the injection spectra, and the fittings are much better than the BPL form, we take the results from the NPI fitting as benchmark. The helium spectrum is the softest at low energies and the hardest at high energies. The Ne spectrum is the hardest in the low-energy range among all species. For C and O, their injection spectra show prominent bumps in the $1-10$ GV range compared with other nuclei. The Fe spectrum is similar to that of Si at low energies, but is slightly harder above 10 GV.

\begin{figure}
\centering
\includegraphics[width=0.48\textwidth]{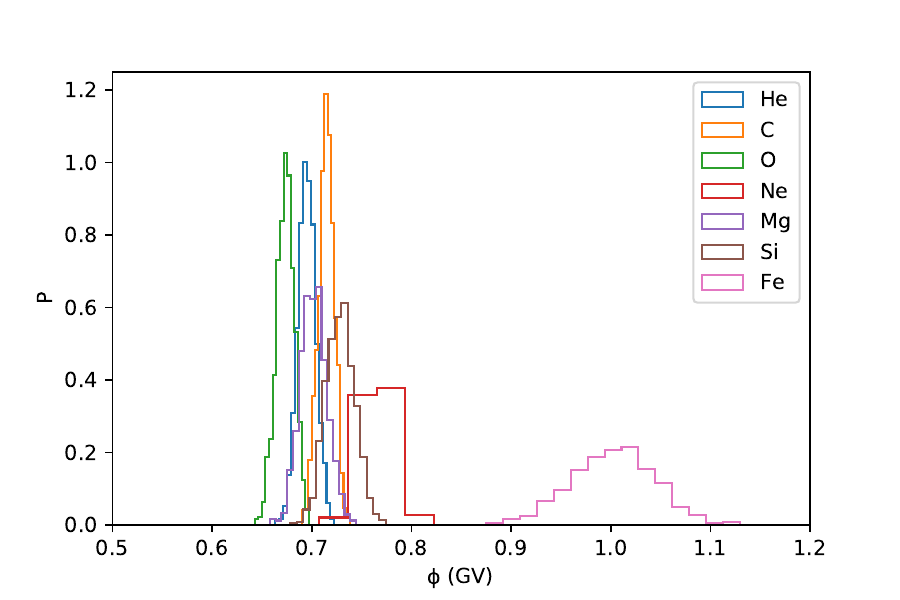}
\includegraphics[width=0.48\textwidth]{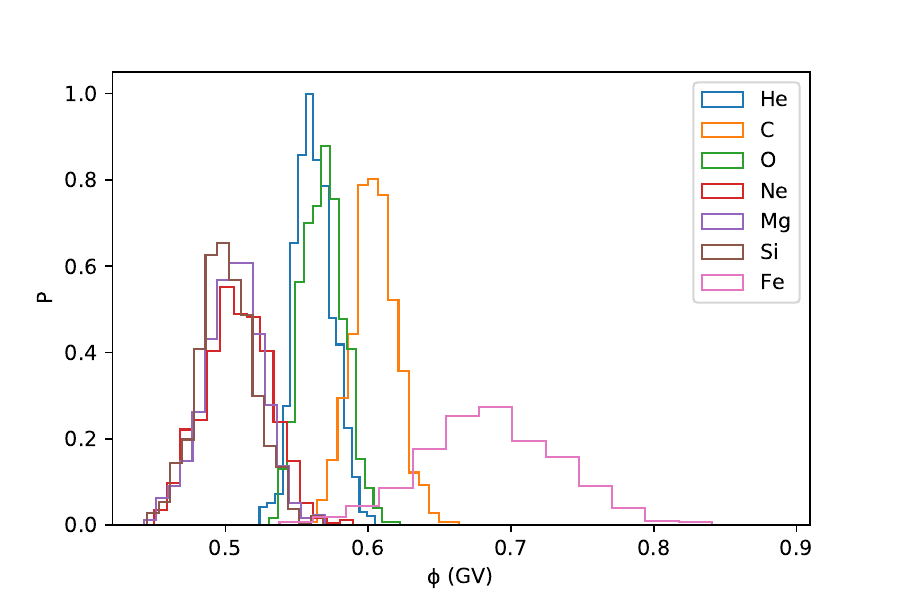}
\caption{The one-dimensional probability density distributions of the 
solar modulation potentials, for the BPL form (left) and NPI form (right).}
\label{fig:phi}
\end{figure}

Thanks to the observations by Voyager-1 outside the solar system and the low-energy 
fluxes by ACE-CRIS, the degeneracy between the injection spectrum and the solar 
modulation can be effectively broken. Fig.~\ref{fig:phi} shows the probability 
distributions of the solar modulation potentials $\phi$ for different nuclei. 
For both the BPL and NPI forms, Fe has the largest $\phi$ value.
The remaining nuclei exhibit similar modulation potentials within $\sim2\sigma$ 
uncertainties. The difference of modulation potentials between Fe and the other
nuclei may be due to the low-energy structures of the Fe spectrum as revealed by
combining the ACE-CRIS and AMS-02 data (see the discussion below).

\begin{figure*}
\centering
\includegraphics[width=0.48\textwidth]{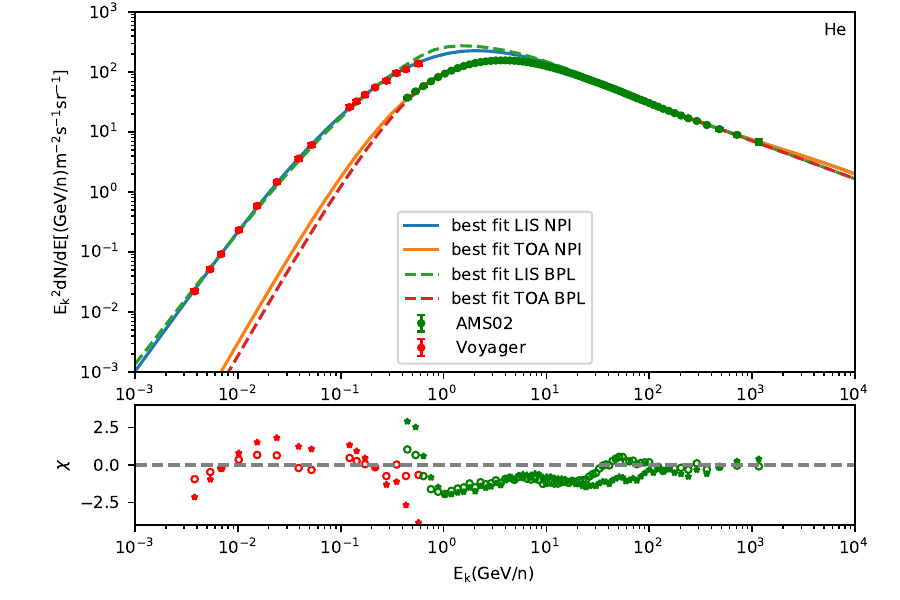}
\includegraphics[width=0.48\textwidth]{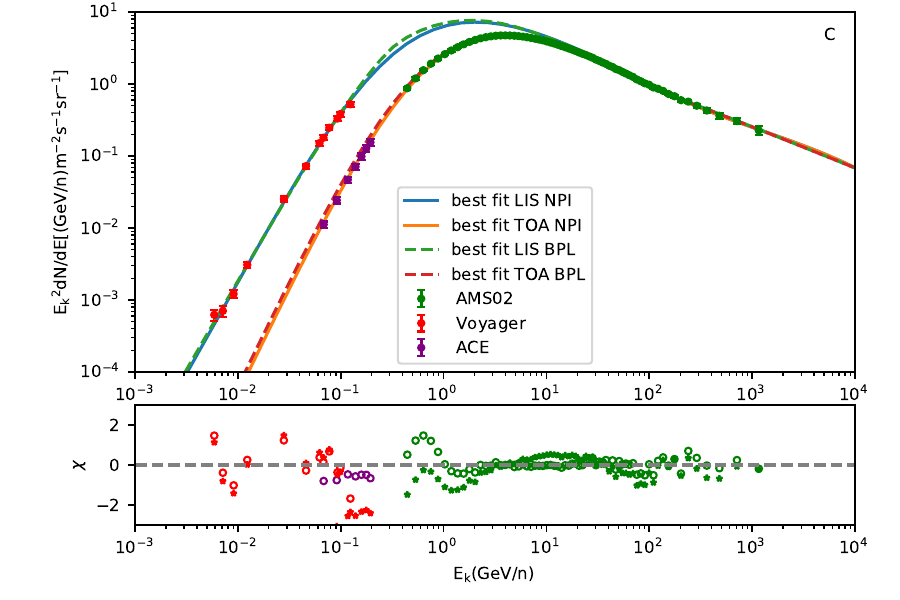}
\includegraphics[width=0.48\textwidth]{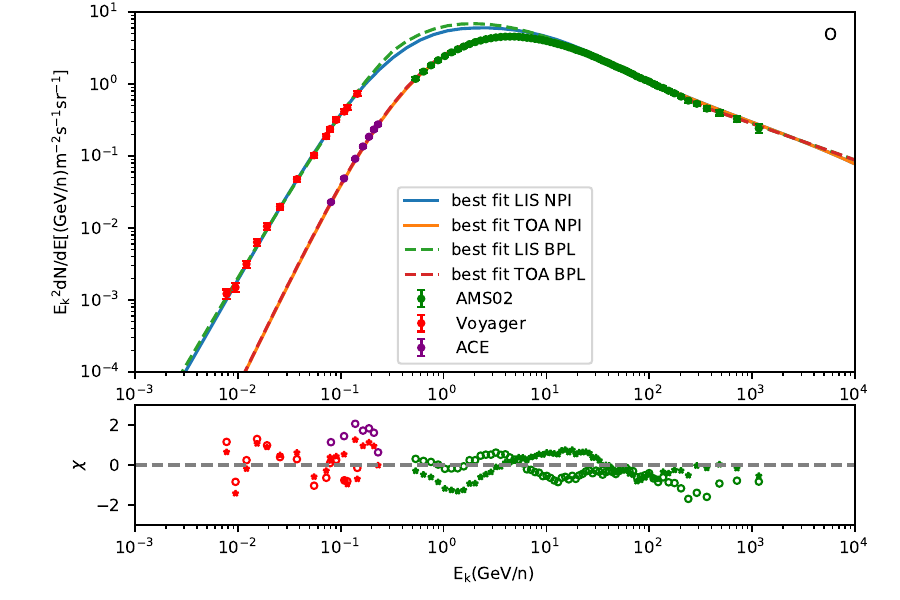}
\includegraphics[width=0.48\textwidth]{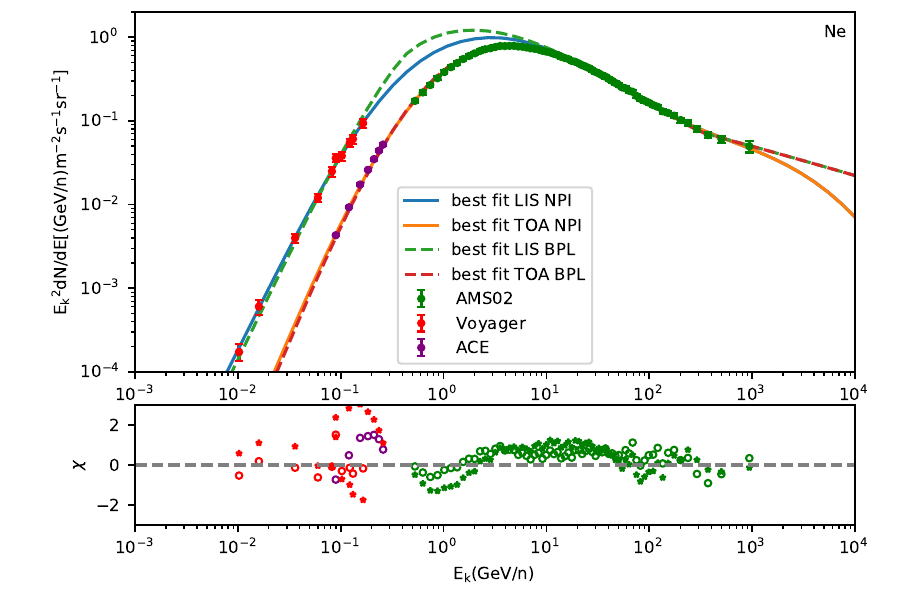}
\includegraphics[width=0.48\textwidth]{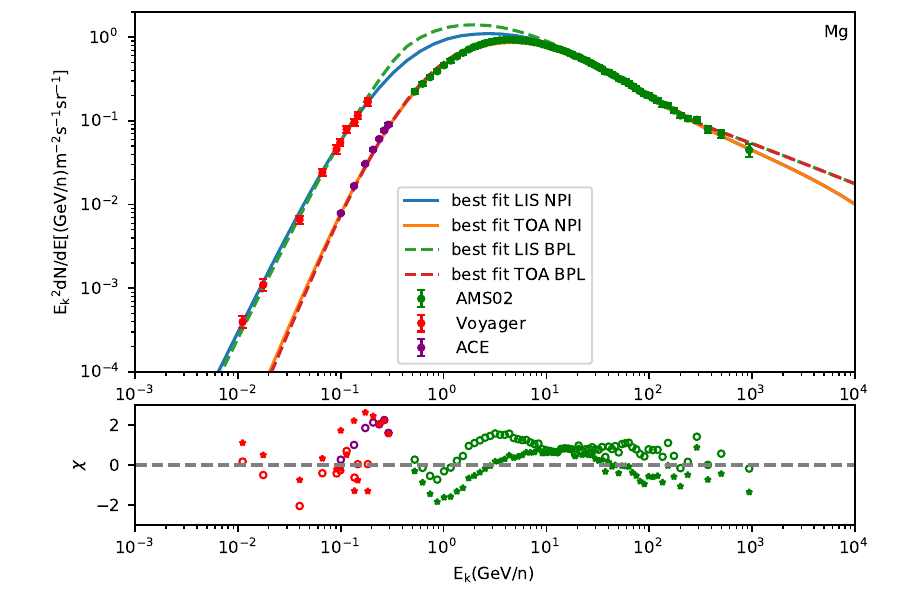}
\includegraphics[width=0.48\textwidth]{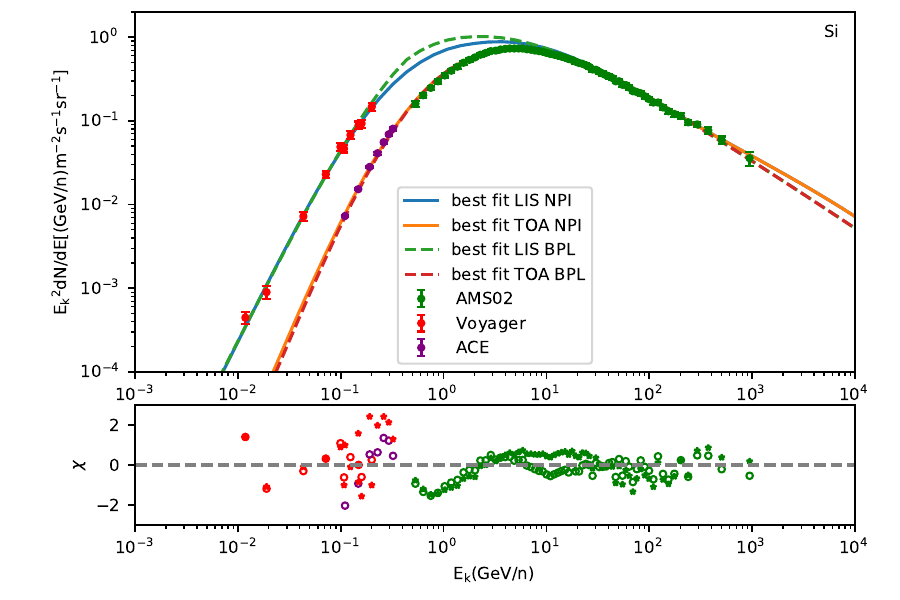}
\includegraphics[width=0.48\textwidth]{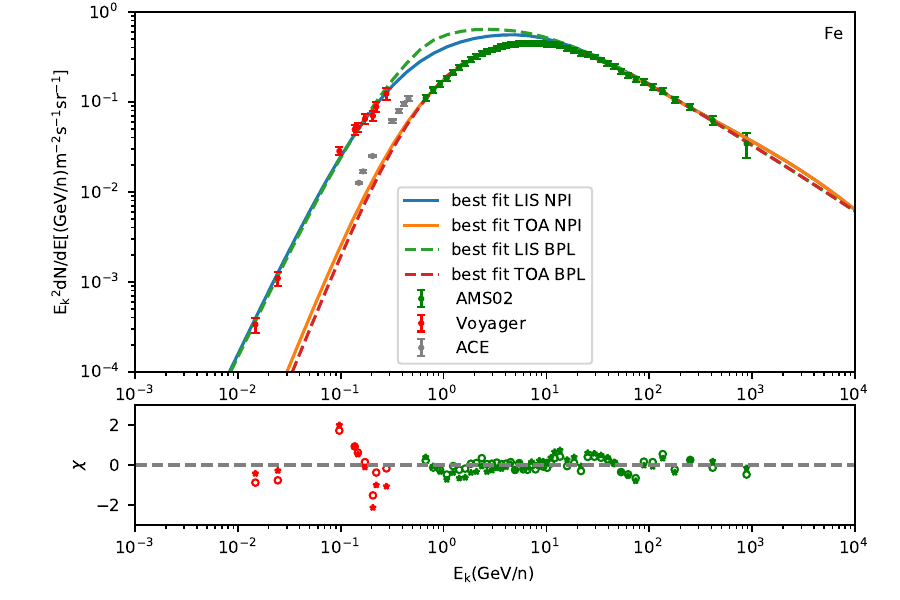}
\caption{Comparison of the best-fit results of the spectra with the 
measurements \citep{2016ApJ...831...18C,2021PhR...894....1A}. In each panel the higher line is the LIS and the lower one is the TOA spectrum. The residuals are depicted in the lower sub-panel, defined as $\chi=({\rm data}-{\rm model})/{\rm error}$ (stars are for BPL form and circles are for the NPI form). 
} 
\label{fig:fitvsdata}
\end{figure*}

Fig.~\ref{fig:fitvsdata} shows the comparisons between the best-fit spectra
and the measurements. The higher curve in each panel represents the local
interstellar spectrum (LIS) before the solar modulation, and the lower one
shows the spectrum at the top of atmosphere (TOA) of the Earth. 
Good consistency between the fitting results and the data can be seen. 
We also show that the ACE-CRIS measurement for the fluxes of Fe nuclei can 
not connect smoothly with the AMS-02 data, as already being pointed out in
\cite{2021ApJ...913....5B}. The combined AMS-02 and ACE-CRIS data may
indicate a bump structure at $\sim2$ GV, which may be due to a past supernova 
explosion in the local bubble \citep{2021ApJ...913....5B}.

\section{Conclusion and discussion}

New measurements of the energy spectra of CRs with precent-level precision
enable us to investigate crucially the acceleration and propagation processes
of particles. The measured spectra contain mixed effects of the acceleration
and propagation, and thus cannot be directly used to infer the injection
spectra of different CR particles. In this work, we thus derive the source 
injection spectra of a series of primary nuclei under the framework of a 
physical propagation model. Our results show that, even these nuclei have
similar $A/Z\approx2$, their injection spectra show diverse behaviors.
As a test, we assume identical injection spectra for all 
these nuclei using the NPI form, and find a reduced chi-squared value of 
$\chi^{2}$/d.o.f.~$=9424.3/586$. If we choose two injection spectra, one is 
applied to He, C, and O nuclei, and the other is applied to Ne, Mg, and Si 
nuclei, we get $\chi^{2}$/d.o.f.~$=650.3/275$ for the fitting to He, C, and O, 
and $\chi^{2}$/d.o.f.~$=223.5/265$ for the fitting to Ne, Mg, and Si. 
The fitting to C and O gives $\chi^{2}$/d.o.f~$=71.2/165$, indicating that
the injection spectrum of C and O should be different from that of He.
When we add Fe to Ne-Mg-Si group, we get $\chi^{2}$/d.o.f.~$=672.0/319$. 
The injection spectrum of Fe appears to be similar to those of Mg and Si 
at low energies, but is harder at high energies. 
The combined fitting of Fe and He-C-O gives $\chi^{2}$/d.o.f.~$=1164.8/339$. 
These tests show that we can perhaps classify the injection spectra into 
four groups, He, C-O, Ne-Mg-Si, and Fe as shown in Fig.~\ref{fig:injection}.
Assuming the same injection spectra for different groups results in poor fittings 
to the data, indicating the intrinsic difference of their injection spectra.

The diversity of the derived injection spectra may be related with the
acceleration processes. Various acceleration models were proposed to explain
the spectral differences of protons and helium nuclei. For example, it was 
proposed that the reverse shock acceleration of different supernova shocks 
(e.g., Type I where hydrogen is absent, and Type II where hydrogen is abundant) 
could explain the harder spectrum of helium nuclei \citep{2013ApJ...763...47P}.
\cite{2011ApJ...729L..13O} proposed that the acceleration in chemically 
enriched regions with outward-decreasing abundance could naturally result in 
different spectra of different species. Those models may be extended to account 
for the differences of injection spectra of heavy nuclei as found in this work.
In addition, models including different ionization histories of nuclei
\citep{1978ApJ...221..703C} and condensation of different elements into grains
\citep{1997ApJ...487..197E} could also explain the diversity of the inferred
injection spectra.

Note that we have assumed a single source population in derive the injection spectra. 
The results may reflect the fact that there are multiple source components of CRs. 
For example, it has been proposed that a nearby source with element abundance 
different from that of the average background sources may result in different 
spectral shapes of various nuclei \citep{2021FrPhy..1624501Y}.

Finally, we assume a spatially uniform propagation in this work. However,
a number of new observations may suggest a spatially-dependent propagation
model of CRs \citep{2012ApJ...752L..13T,2018ApJ...869..176L,
2018PhRvD..97f3008G,2021PhRvD.104l3001Z}. Due to the differences of the
energy loss rates and fragmentation cross sections of different nuclei,
they experience different propagation lengths in the Milky Way. In the
spatially-dependent propagation model, such an effect results in additional
spectral differences on the results based on homogeneous propagation
assumption. Whether or not the observed spectral differences can be
reproduced in a realistic spatially-dependent propagation model needs
future studies.

\normalem
\begin{acknowledgements}
We acknowledge the use of the ACE-CRIS data provided by the ACE Science Center. 
This work is supported by the National Key Research and Development Program of China
(No. 2021YFA0718404), the National Natural Science Foundation of China (No. 12220101003)
and the Project for Young Scientists in Basic Research of Chinese Academy of Sciences 
(No. YSBR-061).
\end{acknowledgements}

\bibliographystyle{raa}
\bibliography{injection}

\end{document}